\documentclass[11pt,twoside,a4paper,cr,final]{cms-tdr}
\def\svnVersion{146080M}

\begin{document}

\hyphenation{had-ron-i-za-tion}
\hyphenation{cal-or-i-me-ter}
\hyphenation{de-vices}

\RCS$Revision: 138024 $
\RCS$HeadURL: svn+ssh://svn.cern.ch/reps/tdr2/papers/XXX-08-000/trunk/XXX-08-000.tex $
\RCS$Id: XXX-08-000.tex 138024 2012-07-19 04:04:00Z alverson $
\newlength\cmsFigWidth
\ifthenelse{\boolean{cms@external}}{\setlength\cmsFigWidth{0.85\columnwidth}}{\setlength\cmsFigWidth{0.4\textwidth}}
\ifthenelse{\boolean{cms@external}}{\providecommand{\cmsLeft}{top}}{\providecommand{\cmsLeft}{left}}
\ifthenelse{\boolean{cms@external}}{\providecommand{\cmsRight}{bottom}}{\providecommand{\cmsRight}{right}}
\cmsNoteHeader{-2012/212} 
\title{
Measurement of charmonium production in PbPb \\
collisions at $\sqrt{s_{NN}}$ = 2.76 TeV with CMS
}

\author{Dong Ho Moon on behalf of the CMS Collaboration}

\address{Korea University, Seoul, Republic of Korea}

\date{\today}

\abstract{
The Compact Muon Solenoid (CMS) is fully equipped to measure hard probes in the di-muon decay channel
in the high multiplicity environment of nucleus-nucleus collisions. Such 
probes are especially relevant for studying the quark-gluon plasma since they are produced
at early times and propagate through the medium, mapping its evolution. CMS has 
measured the nuclear modification factors of non-prompt $J/\psi$ (from b-hadron decays) and
prompt $J/\psi$ in PbPb collisions at $\sqrt{s_{NN}} = 2.76$ TeV. For prompt $J/\psi$ with
relatively high $p_{\rm T}$ ($p_{\rm T}$=6.5-30 GeV/c), a strong, centrality-dependent suppression is
observed in PbPb collisions, compared to the yield in pp collisions scaled by the number of
inelastic nucleon-nucleon collisions. In the same kinematic range, a suppression of non-prompt $J/\psi$,
which is sensitive to the in-medium b-quark energy loss, is measured for the
first time. Results from the 2010 data taking period are reported and an outlook on the 2011 data analysis will be given.
}

\hypersetup{%
pdfauthor={CMS Collaboration},%
pdftitle={
Measurement of charmonium production in PbPb collisions \\
at 2.76 TeV with CMS
}
pdfsubject={CMS},%
pdfkeywords={Quarkonia, suppression, $J/\psi$, prompt, non-prompt, CMS, LHC}}

\maketitle 
\section{Introduction}
\label{Intro}

The strongly interacting matter consists of a deconfined and chirally-symmetric system
of quarks and gluons, called "Quark-Gluon-Plasma" (QGP)
~at large energy densities
and high temperatures. One of the strongest signatures of the existence of the QGP is 
the quarkonia states suppression, which is thought to be a direct effect of deconfinement,
when the binding potential between the constituents of a quarkonium state is screened by the
colour charges of the surrounding light quarks and gluons. However, there are further possible
changes to the quarkonium production in heavy-ion collisions. One initial effect, i.e.  
the modifications of the parton distribution functions inside the nucleus (nuclear shadowing),
can reduce the production of quarkonia without the presence of a QGP~\cite{QGP2}. A second effect, 
related to the large number of heavy quarks produced in heavy-ion collisions in particular at the energies
accessible by the Large Hadron Collider (LHC), could lead to an increased production of quarkonia
via statistical recombination~\cite{QGP5}.

The suppression (or enhancement) of $J/\psi$ can be quantified by the ratio of produced $J/\psi$ in AA collisions to those in pp collisions 
scaled by the number of binary nucleon-nucleon collisions ($N_{\rm coll}$), 
known as the nuclear modification factor $R_{\rm AA}$:

\begin{equation}
R_{\rm AA} = \frac{\mathcal{L}_{\rm pp}}{T_{\rm AA} N_{\rm MB}}\frac{N_{\rm PbPb} (J/\psi)}{N_{\rm pp} (J/\psi)} \cdot \frac{\varepsilon_{\rm pp}}{\varepsilon_{\rm PbPb}}
\label{eqn:def_raa}
\end{equation}

where $T_{\rm AA}$ = $<N_{\rm coll}>/\sigma^{\rm NN}_{\rm inel}$ can be calculated from a Glauber model accounting for the nuclear collision
geometry~\cite{Glauber}. $\mathcal{L}_{\rm pp}$ is the integrated luminosity of pp collisions and $N_{\rm pp}$ or $N_{\rm PbPb} (J/\psi)$ are the raw yields 
of $J/\psi$ measured in pp and PbPb collisions. 
$N_{\rm MB}$ is the count of minimum bias events in PbPb, and $\varepsilon_{\rm pp}/\varepsilon_{\rm PbPb}$ is the multiplicity dependent ratio of the efficiencies
in pp and PbPb collisions for trigger and reconstruction, which is determined by a Monte-Carlo simulation that embedded a PYTHIA signal event~\cite{Tsj} to a HYDJET
background event~\cite{IpLok}. In the simulation, $\varepsilon_{\rm pp}/\varepsilon_{\rm PbPb}$ was estimated to be about 1.17 for the most central bin.
The results are cross checked with real data using a tag-and-probe technique~\cite{TnP}.

\section{Data samples and analysis procedure}
\label{DataAna}

The measurement presented here is based on $\sqrt{s_{_{NN}}}=2.76$~TeV PbPb data samples corresponding to integrated luminosities of 7~$\mu b^{-1}$ and 150~$\mu b^{-1}$, collected by the CMS experiment in 2010 and 2011, respectively. The pp reference measured at the same nucleon-nucleon collision energy corresponds to an integrated luminosity of 230~$\mu b^{-1}$.

The central feature of the CMS apparatus is a superconducting solenoid with 3.8~T magnetic field.
Immersed in the magnetic field are the silicon pixel and strip
tracker, the lead-tungstate crystal electromagnetic calorimeter, and the brass/scintillator hadron calorimeter. 
Muons are measured in gas ionization detectors embedded in the steel return yoke and
in the pseudorapidity window $|\eta| <$ 2.4, with detection planes made of three technologies:
Drift Tubes, Cathode Strip Chambers, and Resistive Plate Chambers. 
A more detailed description of the CMS detector can be found in Ref.~\cite{JINST}.

In addition, CMS classified the measured $J/\psi$ into the prompt component associated with the primary vertex and the 
non-prompt component associated with a secondary vertex of $\mu^{+}\mu^{-}$\cite{CMSJpsi2011}. 
Because B-mesons fly finite pathlength before they decay, the non-prompt $J/\psi$ can be separated from
the prompt ones by the pseudo-proper decay length:

\begin{figure*}
    \centering
    \includegraphics[width=0.4\textwidth]{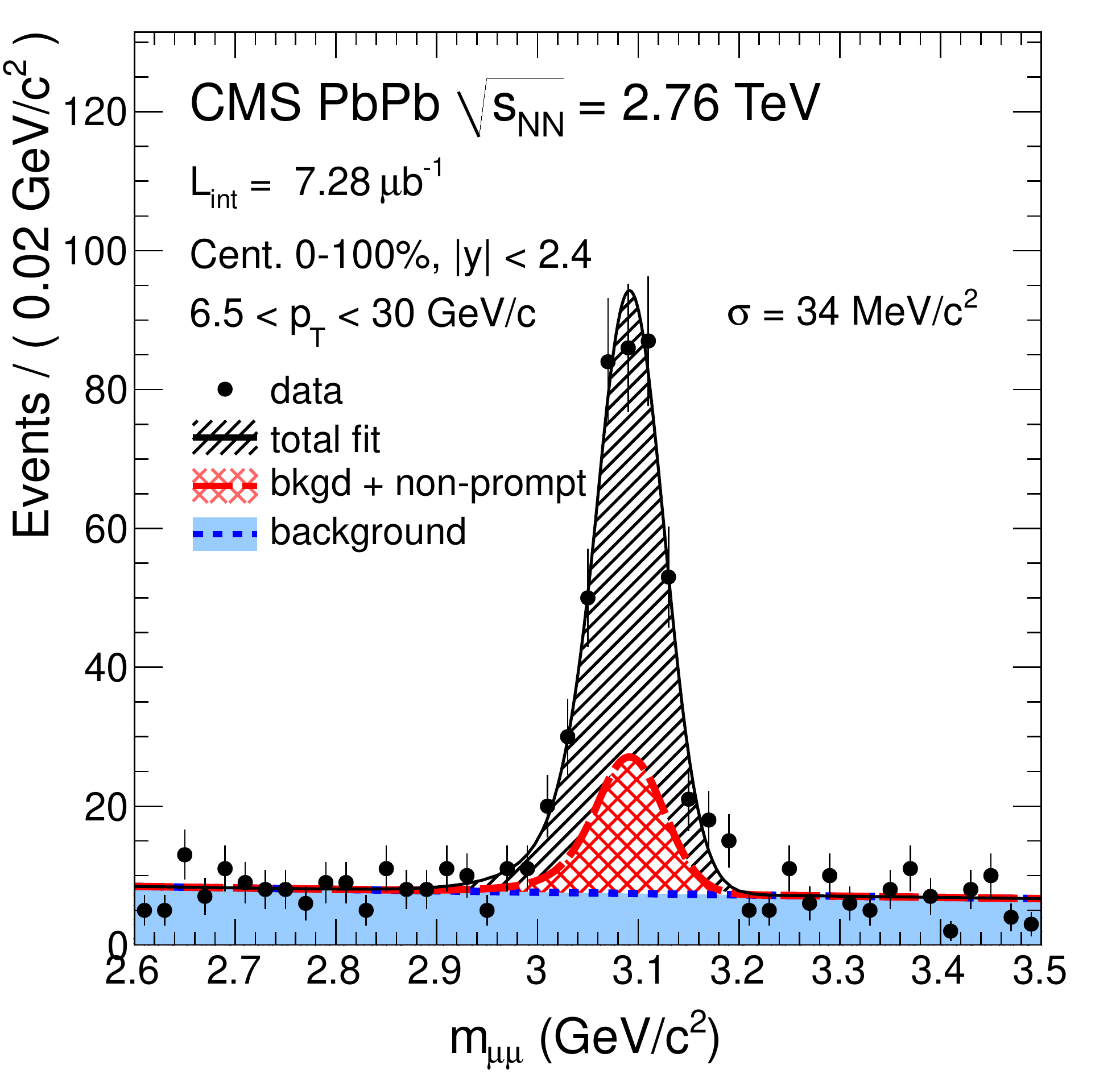}
    \includegraphics[width=0.4\textwidth]{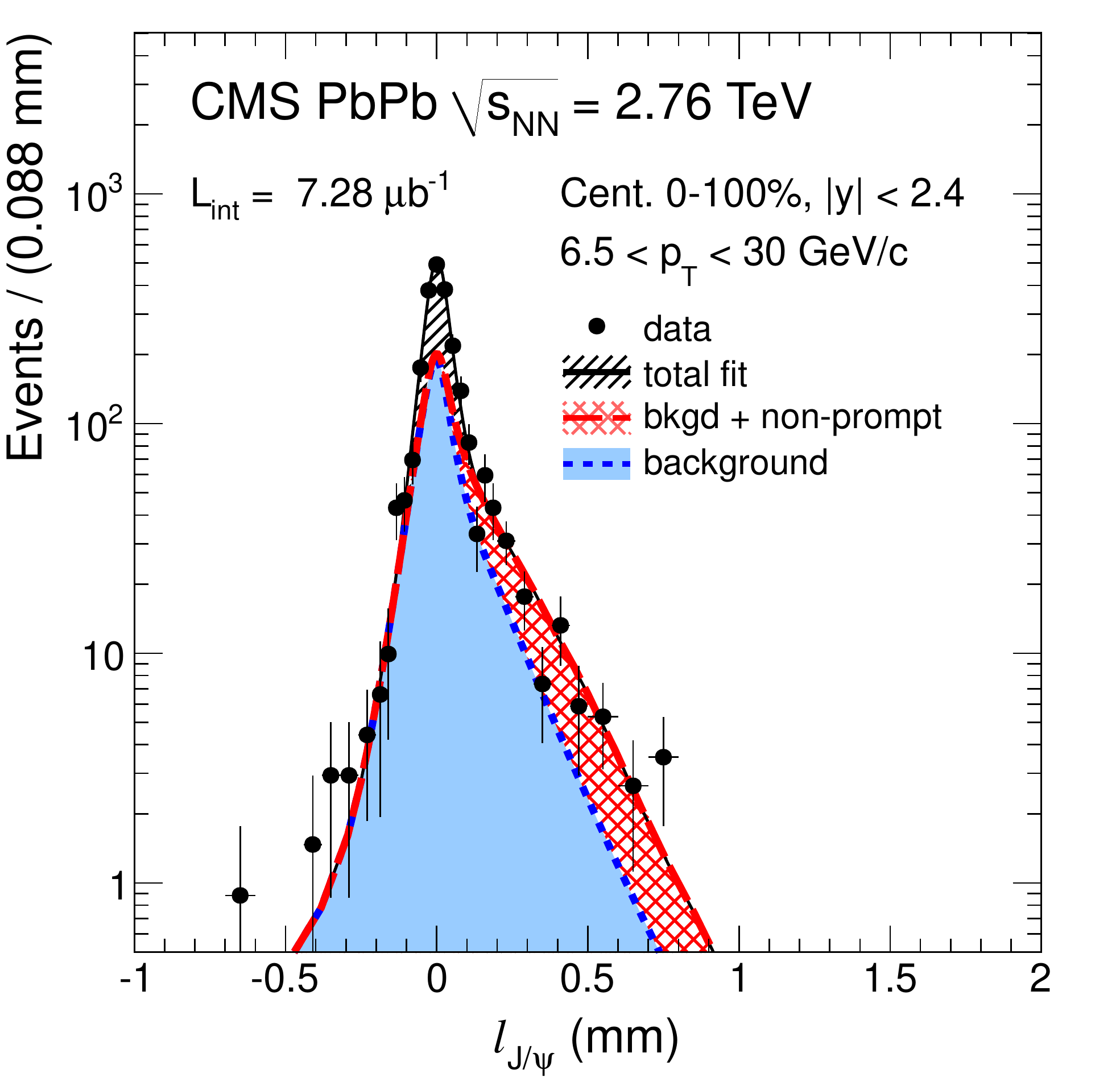}
    \caption{Invariant mass (left) and pseudo-proper decay length (right) distributions of $\mu^{+}\mu^{-}$ pairs for $|y| < 2.4$ in minimum bias 
             PbPb collisions at $\sqrt{s_{\rm NN}} = 2.76$ TeV from CMS \cite{CMSJpsi2011}. The data (black circles) are overlaid with the projections of
             the 2-dimensional fit. The different contributions are background (dotted blue line), non-prompt $J/\psi$ (dashed red line), and sum
             of background, non-prompt and prompt $J/\psi$ (solid black line).
      }
    \label{fig:SigExt}
\end{figure*}

\begin{equation}
l_{J/\psi} = L_{xy} \frac{m_{J/\psi}}{p_{\rm T}}
\label{eqn:Ljpsi}
\end{equation}
where $L_{xy}$ is the distance between the primary and the secondary vertices in the transverse plane
and $m_{J/\psi}$ is the rest mass of $J/\psi$. 
Fig.\ref{fig:SigExt} shows the invariant mass and the projected $l_{J/\psi}$ distributions of $\mu^{+}\mu^{-}$ pairs in PbPb
collisions at $\sqrt{s_{\rm NN}} = 2.76$ TeV \cite{CMSJpsi2011}.

\begin{figure*}
    \centering
    \includegraphics[width=0.4\textwidth]{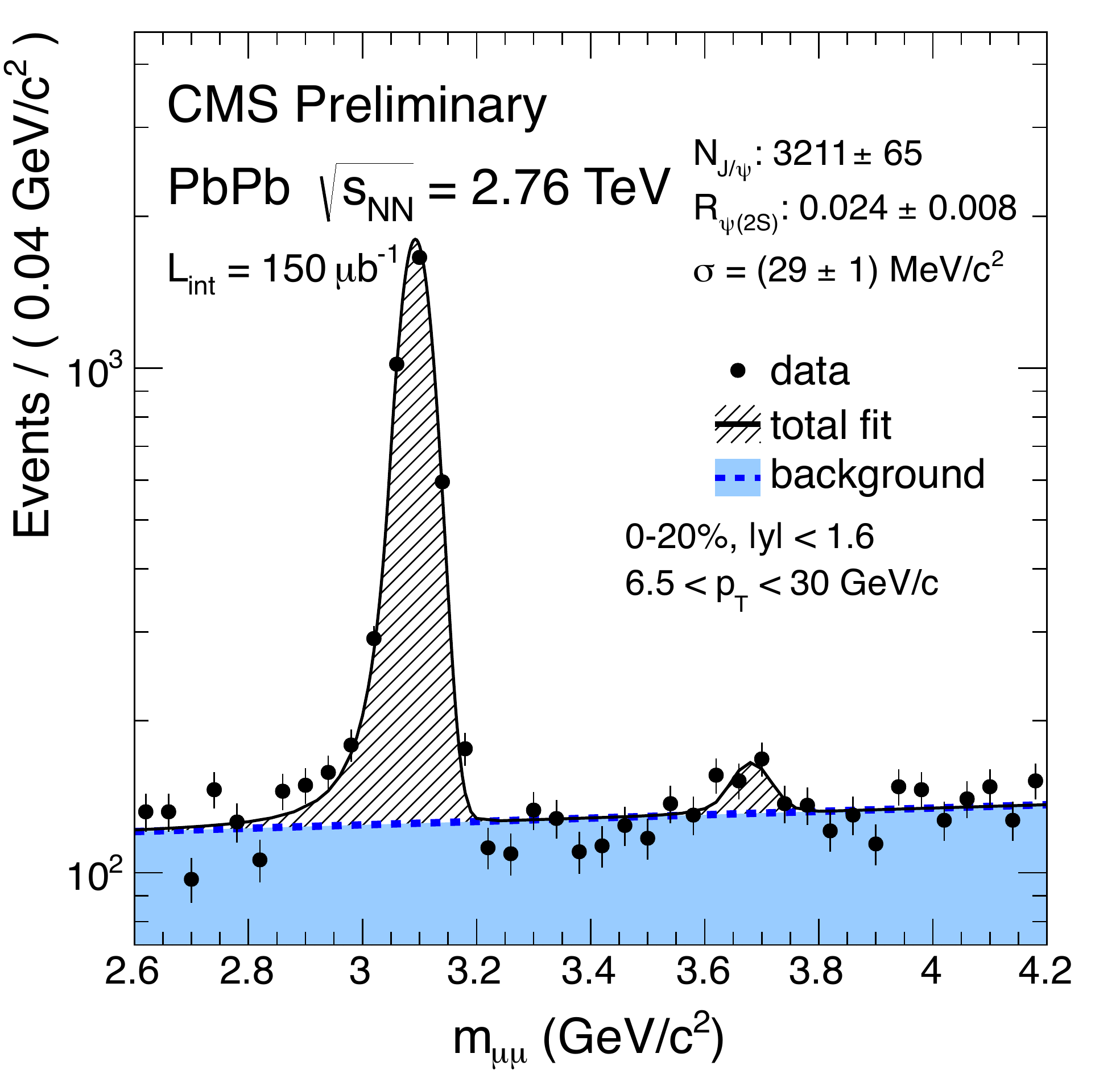}
    \includegraphics[width=0.4\textwidth]{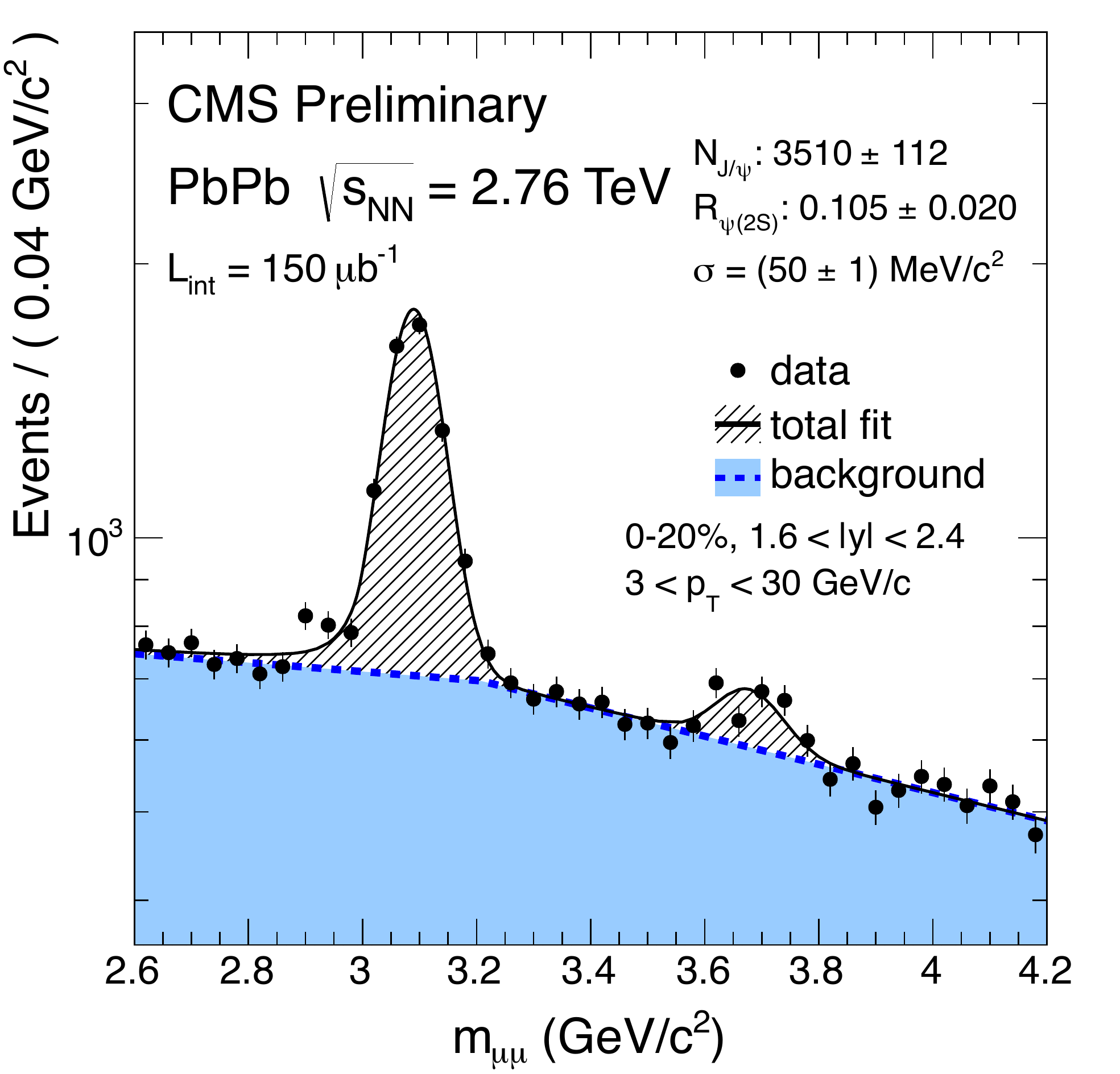}
    \caption{
             Invariant-mass spectrum of $\mu^{+}\mu^{-}$ pairs (black circles) in  
             0–-20$\%$ most central PbPb collisions. The $p_{\rm T}$ and rapidity bins are (left) 6.5 $< p_{\rm T} <$ 30 GeV/c and $|y| <$ 1.6,
             (right) 3 $< p_{\rm T} <$ 30 GeV/c and 1.6 $< |y| <$ 2.4. 
             The fit to the data with the functions (CrystalBall+Exponential) is shown as the black line. The
             dashed blue line shows the fitted background contribution~\cite{PsiPAS}. 
             $R_{\psi(2S)}$ is the ratio for the raw yield of $J/\psi$ and $\psi(2S)$.
             }         
    \label{fig:MassPsi}
\end{figure*}

The huge amount of 2011 data, 20 times larger than 2010, with 150 $\mu b^{-1}$ also allows the measurement of $\psi$(2S) in CMS.
Fig.\ref{fig:MassPsi} shows the invariant-mass distribution of $\mu^{+}\mu^{-}$ pairs in the region 
2.6 $< m_{\mu^{+}\mu^{-}} <$ 4.2 GeV/$c^{2}$ for PbPb collisions, for the following two kinematic ranges~\cite{PsiPAS}:
\begin{itemize}
\item Lower-$p_{\rm T}$ and forward rapidity : 3 $< p_{\rm T} <$ 30 GeV/c, 1.6 $< |y| <$ 2.4 
\item Higher-$p_{\rm T}$ and mid-rapidity : 6.5 $< p_{\rm T} <$ 30 GeV/c, $|y| <$ 1.6 
\end{itemize}

\section{Results}

\begin{figure*}
    \centering
    \includegraphics[width=0.4\textwidth]{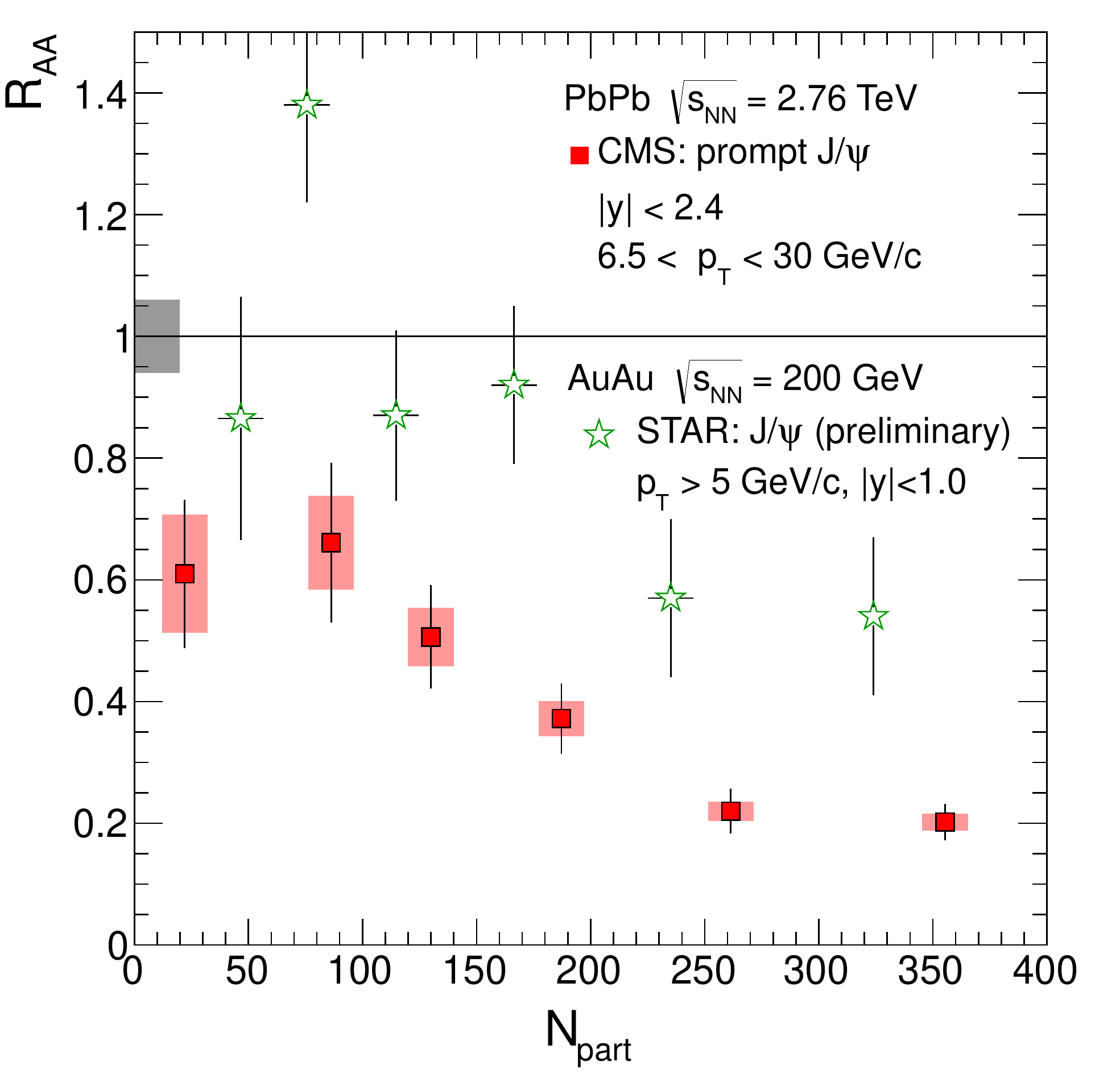}
    \includegraphics[width=0.4\textwidth]{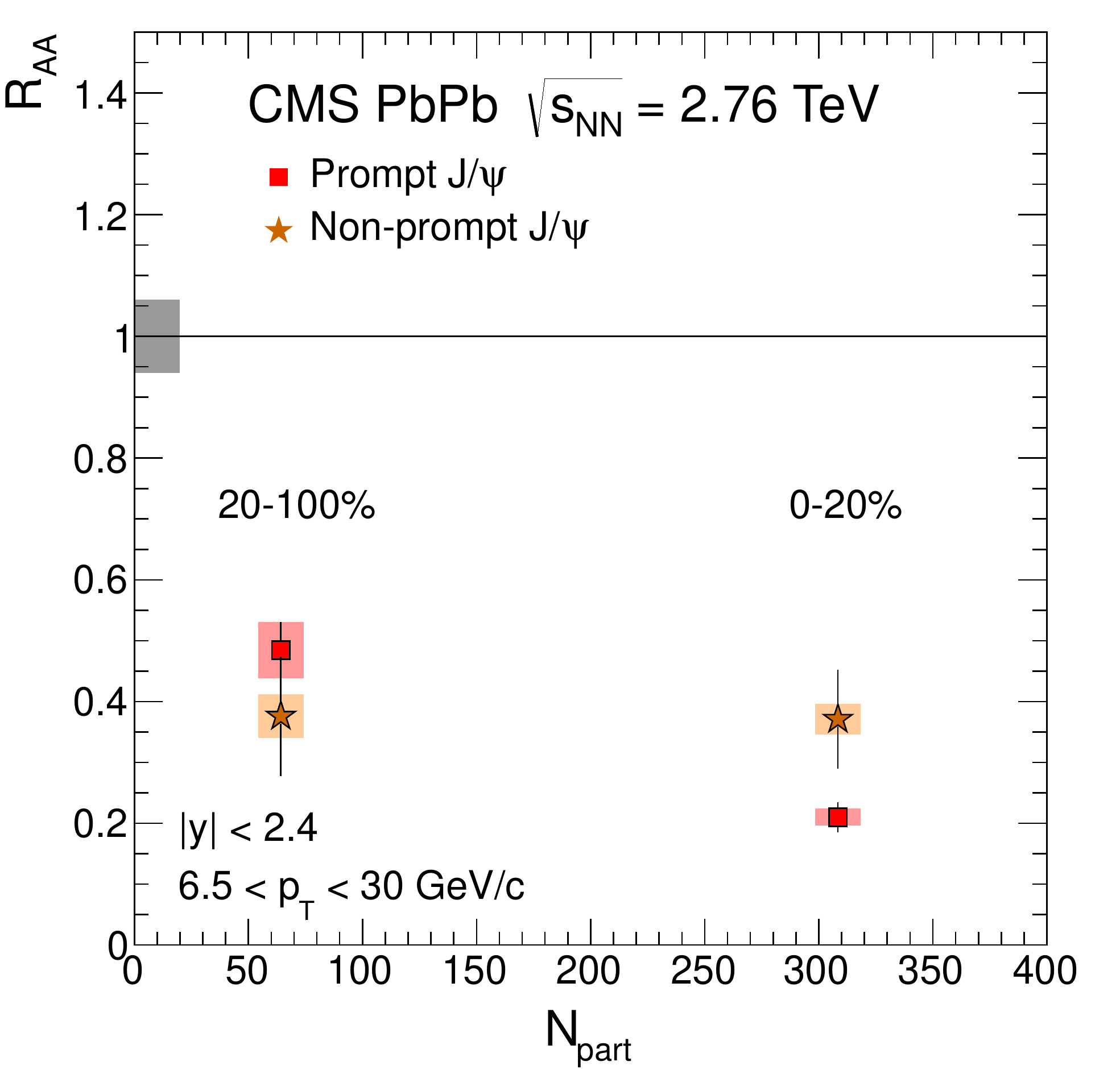}
    \caption{$R_{\rm AA}$ of prompt (red squares) and non-prompt $J/\psi$ (orange stars) as a function of $N_{\rm part}$~\cite{CMSJpsi2011, RaaSTAR}. 
             The data of prompt $J/\psi$ are compared to STAR (green stars) results in the left panel. 
             A global uncertainty of 6$\%$, from the integrated luminosity of the pp data sample, is shown as a grey box at $R_{\rm AA}$ = 1. 
             Statistical (systematic) uncertainties are shown as bars (boxes)~\cite{CMSJpsi2011}.
      }
    \label{fig:RaaNpart}
\end{figure*}

Fig.\ref{fig:RaaNpart} presents $R_{\rm AA}$ of prompt and non-prompt $J/\psi$ (from B-meson decay) 
with 6.5 $< p_{\rm T} <$ 30 GeV/c and $|y| <$ 2.4 as a function of the number of participating nucleons ($N_{\rm part}$). 
A significant centrality dependent suppression of prompt $J/\psi$ production is observed.
In the 10$\%$ most central events, the suppression reaches a factor of 5 with respect to the pp reference. The suppression decreases towards peripheral events to a
factor $1.6$ in the 50$-$100$\%$ centrality bin. The results are compared to $R_{\rm AA}$ of $J/\psi$ measured in AuAu collisions at $\sqrt{s_{\rm NN}} = 200$ GeV 
by STAR~\cite{RaaSTAR}
, which measured less suppression than CMS. The $R_{\rm AA}$ of non-prompt $J/\psi$ has been also measured, which gives some hint of in-medium
energy loss of b-quarks~\cite{CMSJpsi2011} in the right panel of Fig.\ref{fig:RaaNpart}.

\begin{figure*}
    \centering
    \includegraphics[width=0.8\textwidth]{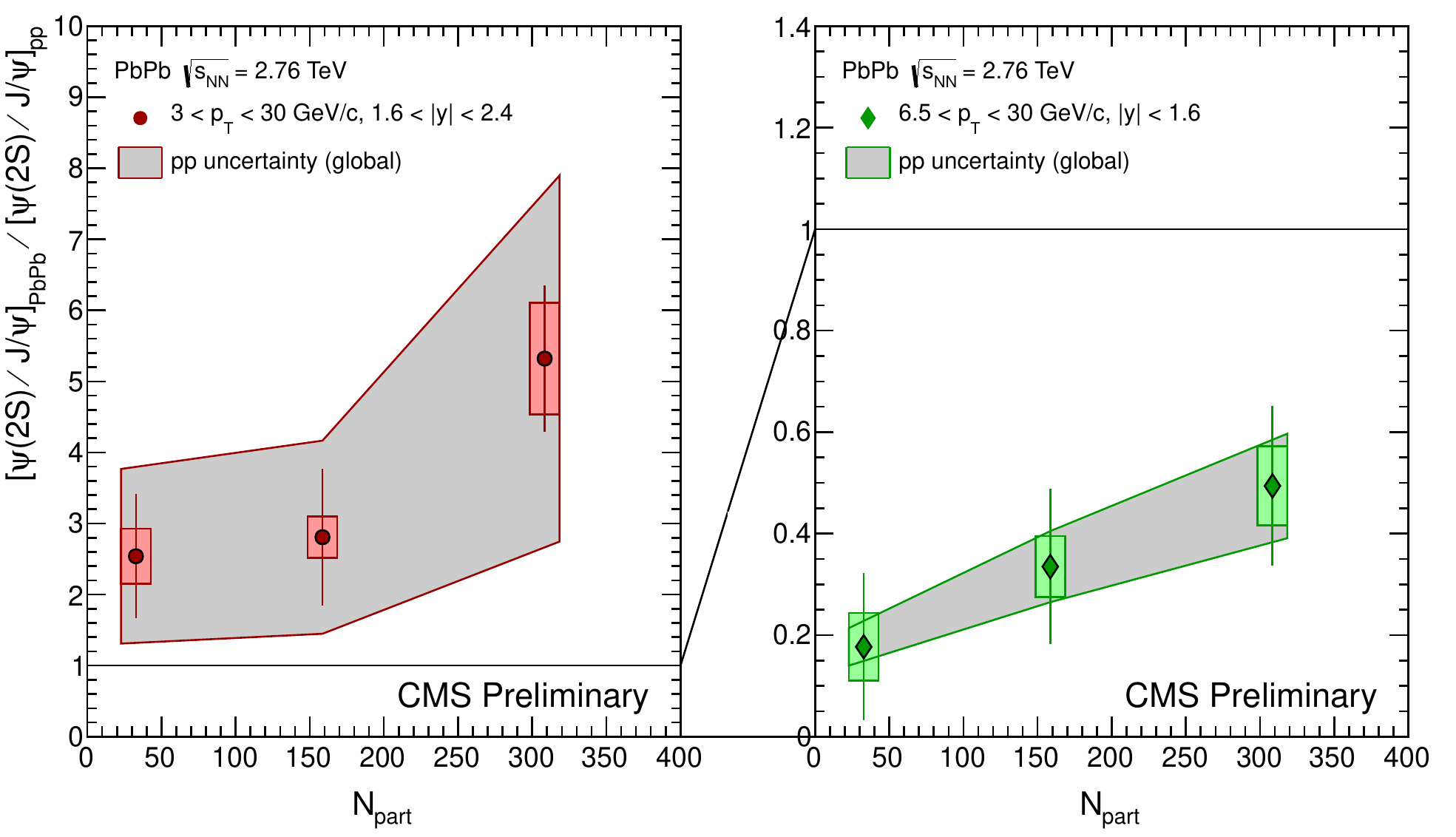}
    \caption{Measured yield double ratio $(N_{\psi(2S)}/N_{J/\psi})_{\rm PbPb} / (N_{\psi(2S)}/N_{J/\psi})_{\rm pp}$ as a function of centrality. 
             The $p_{\rm T}$ and rapidity bins are 6.5 $< p_{\rm T} <$ 30 GeV/c and $|y| <$1.6 (right) and 3 $< p_{\rm T} <$ 30 GeV/c, 1.6 $< |y| <$ 2.4 (left).
             The error bars and boxes stand for the PbPb statistical and systematic uncertainties, respectively. The shaded band is the uncertainty on the pp
             measurement, common to all double-ratio points~\cite{PsiPAS}.
      }
    \label{fig:DoubleRatio}
\end{figure*}

As for $\psi$(2S) measurements, the results are presented as the form of double ratio, the ratio of the single ratio of
$J/\psi$ and $\psi$(2S) measured in pp and PbPb, shown in Fig.\ref{fig:DoubleRatio} 
as a function of $N_{\rm part}$ in both the lower $p_{\rm T}$ (left) and higher $p_{\rm T}$ (right) selections.
The double ratio measured in $p_{\rm T} >$ 6.5 GeV/c and $|y| <$ 1.6
is always less than unity, which means that high-$p_{\rm T}$ $\psi$(2S) are more suppressed than
$J/\psi$. Within uncertainties, no significant centrality dependence is observed.
However, in the lower $p_{\rm T}$ range (3 $< p_{\rm T} < $ 30 GeV/c) and at the forward rapidity, 
the data show an increase of the double ratio with
centrality and the double ratio is larger than unity, though with large uncertainties.
In the most central collisions the double ratio is
$5.32\pm1.03\rm{(stat.)}\pm0.79\rm{(syst.)}\pm2.58\rm{(pp)}$, meaning that more $\psi$(2S) are produced compared
to $J/\psi$ than in pp collisions, again with large uncertainties, which came from the insufficient pp statistics~\cite{PsiPAS}.

\section{Acknowledgements}
\label{Ackn}
The author wishes to thank the Korea CMS for their support.

\end{document}